\documentclass[aps,amsart
amsproc,article,elsarticle,emulateapj,superscriptaddress]{revtex4}
\usepackage{bm}
\usepackage{graphicx,amsmath,latexsym,amssymb}
\usepackage{xcolor}
\usepackage{sistyle}
\usepackage{pstricks,pst-grad}

\usepackage{pstricks,pst-grad}
\begin{document}
\title{Quantum localization corrections from the Bethe-Salpeter equation}
\author{Afifa Yedjour}
\affiliation{Facult\'{e} de Physique, USTO MB, B.p.1505 El M'Naour, 31000.Oran, Alg\'{e}rie.}
\author{Abdelaali Boudjemaa*}
\affiliation{ Department of Physics, Faculty of Exact Sciences and Informatics, Hassiba Benbouali University of Chlef, P.O. Box 78, 02000, Chlef, Algeria.}
\email[Electronic address:]{a.boudjemaa@univ-chlef.dz}
\date{\today}

\begin{abstract}  

We investigate coherent matter wave transport in isotropic 3D speckle potentials by using the Bethe-Salper equation and the self-consistent theory of
localization. This model constitutes an efficient tool to properly evaluate corrections to Boltzmann diffusion by taking into consideration quantum interference terms 
between the multiple-scattering paths.
We calculate analytically and numerically the static current density, the density of states, the dipolar contribution and the reduced diffusion coefficient.
Our results reveal that  quantum corrections to diffusive transport, known as weak localization may not only lead to shift the above quantities but 
affect also the position of the mobility edge.
\end{abstract}
\maketitle


\section{Introduction}

It has been established that the transport properties of a quantum particle in a disordered system are intrinsically determined
by the interference of several scattering paths, which can lead to a spacial localization \cite{ref1,ref2,ref3,ref4}.
The quantum particle then remains localized around its initial position, and it leads to a  total suppression of transport, giving rise to arrest 
the diffusion and  hence, cancel the conductivity \cite{ref5,ref6}.

Experimentally, the localization has been observed in many systems including  light waves \cite{Wier,Sche,Sto,Sch,Lah}, microwaves \cite{Dal,Chab}, sound waves \cite{Weav}, 
electron gases \cite{ref24}, and coherent  matter waves \cite{ref7,ref8}. 
Moreover, ultracold atoms open a new scenario for studying disorder-induced localization, due to high degree of  interactions.
Experimental observation of the Anderson transition  of coherent  matter waves in a disordered optical potential has been reported in Refs.\cite{ref7,ref8}. 
In three dimensions (3D), experimental observation of Anderson localization in matter waves of dilute Bose and Fermi gases
in a speckle potential has been reported in \cite{ref10,ref9}.


One could argue that in disordered media the macroscopic transport properties namely diffusion, weak and strong localization, depend on the 
statistical properties of the disorder potential \cite{ref11}.
The most interesting feature in the statistical properties of transport is the return probability to a given point in which all scattering paths are closed loops \cite{ref11,kunn}.
A weak localization comes from the fact that the probability of the wave to return to its initial position is possible through loop paths.
Each loop can generate two multiple scattering paths along which exactly the same phase accumulates in successive scattering events results in  constructive  interferences of the matter wave. 
This coherent effect which is valid for any specific realization of the disorder potential leading to a diffusive transport for which the diffusion coefficient is  reduced.
On the other hand, if the disorder is strong the propagation of a  coherent wave is stopped after a certain time in any dimension indicating that the diffusion coefficient  
vanishes. Therefore, the return probability decreases exponentially from a certain point in space with a characteristic length known as the localization length \cite{ref12}.
Recently, quantum transport and Anderson localization of atomic matter waves in 3D anisotropic disordered potentials have been investigated in \cite{Piraud,Piraud1,ref19}.

In the present paper, we use the Bethe-Salper equation and the self-consistent Born approximation to study the transport properties of ultracold atoms exposed to isotropic 3D speckle potentials
far below the quantum degenerate regime (low densities).
This model helps deal with the transport phenomenon of disordered BEC.
We calculate in particular quantum corrections stemming from the interference effects {\it known as weak localization} to classical transport.
Starting from the quantum kinetic theory, we calculate the current density and determine all relevant quantities such as the spectral function, 
the dipolar contribution, and the reduced diffusion coefficient that are necessary to describe the average diffusion process.
Some useful analytic expressions are obtained in some limits.

Furthermore, we present our numerical solutions of the Bethe-Salpeter equation for the transport phenomenon of BEC subjected to a laser speckle potential. 
By means of the self-consistent Born approximation we evaluate iteratively the self-energy  and the spectral function. 
This allows us to identify numerically the current density, the density of state, the fraction of the localized atoms and the diffusion coefficient.  
The main results emerging from this analysis is that corrections due to the interference effects may alter the overall behavior of these quantities and thus,
affect the coherent transport regime.
We show that the  diffusion constant which depends on the atomic energy and on the disorder amplitude falls to zero due to the weak localization effects.
The weak localization to the diffusion constant  can be measured by releasing a confined atomic cloud and monitor its long-time spread inside the speckle
field by time-of-flight or in-situ imaging techniques \cite{kunn}.
The increasing of the atomic energy leads to modify  the position of the  mobility edge.
The most advantage of our method is that it does not require cutoffs to eliminate short wave paths that diverge even when scattering extends to infinite wave numbers
in contrast to the theories of Refs.\cite{kunn,Piraud}.
Our results provide new insights into the theory of Anderson localization and diffusion of matter wave in isotropic optical disorders.

The rest of the paper is organized as follows.
In section \ref{CTDM} we introduce the basic concepts of quantum transport of matter waves in disordered media and present the Bethe-Salpeter equation and its 
the eigenfunction. 
We obtain also an analytic expression describing the quantum localization corrections for the diffusion constant in the case of an isotropic scattering.
Section \ref{NR} is devoted to the numerical solutions of the Bethe-Salpeter equation for the transport phenomenon of BEC subjected to a laser speckle potential
and compare them with classical regime. The  current density and the reduced diffusion constant are also compued and compared with the analytical results.
In section \ref{Conc} we present a brief summary.


\section{Coherent transport in a disordered medium} \label{CTDM}

We consider a cloud of noninteracting cold atoms that is described by the single particle Hamiltonian 
$H=p^2/2m+V({\bf r})$, where $V({\bf r})$ is a static random potential introduced
after the harmonic confinement in the transport directions has been switched off as in the experiments realized in \cite{ref7,ref8}. 
In such a regime of low densities, the  evolution of the condensate is described by a linear time-dependent Schr\"odinger equation with a random potential. 
This latter is assumed to satisfy the following statistical properties:
$\langle V(\mathbf r)\rangle=0$, and  $\langle V(\mathbf r) V(\mathbf r')\rangle=R (\mathbf r-\mathbf r')$,
where $ \langle \bullet \rangle$ denotes the disorder ensemble average and $R (\mathbf r-\mathbf r')$ is the disorder correlation function. 
In the following we deal with coherent transport of such noninteracting cold atoms in an isotropic disordered environment using the Bethe-Salpeter equation.\\
We set $\hbar=1$ throughout the manuscript.


\subsection{Solutions of Bethe-Salpeter equation} \label{DCBSE}

Let us start by considering two particles of Green function $\phi_{\bf p,p'}(\varepsilon, \Omega,\ {\bf q})$ which describes the probability density  of matter wave in a disorder potential.
We seek for the hydrodynamic expression which is defined as \cite{kunn,Piraud,ref15}:
\begin{equation}\label{equa1}
\phi_{\bf p,p'}(\varepsilon, \Omega,\ {\bf q}) \equiv \overline{\langle {\bf p}_{+}| G(\varepsilon_{+})| {\bf p}'_{+}\rangle
\langle {\bf p}'_{-}| G^{\dagger}(\varepsilon_{-})| {\bf p}_{-}\rangle},
\end{equation}
where $G$ is the retarded Green operator, $\varepsilon_{\pm}= \varepsilon \pm \Omega/2$ , 
${\bf p}_{\pm}= {\bf p \pm q}/2$,
${\bf p'_{\pm}= p' \pm q}/2$ with ($q$ and $\Omega$ are the Fourier conjugate variables of space and time, respectively \cite{ref15}). 
Our target is to analyze the behavior of the density diffusion for large distance and long times.
We anticipate that at long time ($\Omega \rightarrow 0)$ and large distance $q \rightarrow 0$, the probability reads:
\begin{equation}\label{equa2}
 \phi_{\bf p,p'}(\varepsilon, \Omega,\ {\bf q}) = K\frac{L_{\varepsilon, {\bf p}}(\Omega, {\bf q}) L _{\varepsilon, {\bf p'}} (\Omega,{\bf q})}{- i \Omega+ D(\varepsilon) q^{2}},
\end{equation}
where $K$ is a normalized constant which will be determined later,  $L_{\varepsilon, \bf p}(\Omega,{\bf q})$ is the eigenfunction of the Bethe-Selpeter equation associated 
with the hydrodynamic diffusion, and $D(\varepsilon)$ is the diffusion constant. The density diffusion $\phi_{\bf p,p'}(\varepsilon, \Omega,\ {\bf q})$
is controlled by the Bethe-Salpeter equation, which can be formally written as \cite{Piraud,Piraud1}:
\begin{equation}\label{equa3}
 \phi= \overline{G}\otimes \overline{G}^{\dagger}  + \overline{G}\otimes\overline{G}^{\dagger} U \phi.
\end{equation}
The first term in Eq.~(\ref{equa3}) represents the intensity propagation with uncorrelated disordered potential.
The second term accounts for  quantum corrections, involves the vertex function $U_{\bf p,p'}$ which includes  all correlations between different amplitudes in the density propagation.
The Green function is defined as : $ G({\bf p},\varepsilon) = [\varepsilon - \varepsilon_{\bf p}\ -\Sigma(\varepsilon,\ {\bf p})]^{-1}$, where $\varepsilon_{p} =p^{2}/2 m$ 
is the kinetic energy of the particle, and $\Sigma(\varepsilon, {\bf p})$ is the self-energy which has already been examined in \cite{ref22}.

The product of the average propagators  $\overline{G}^{\dagger}(\varepsilon_{-},{\bf p}_{-}) \otimes \overline{G}(\varepsilon_{+},{\bf p}_{+})$ 
can be reformulated in momentum space by using the identity 
$\overline{G}^{*} *\overline{G}= \big(\overline{G}^{*} - \overline{G}\big) /\big(\overline{G}^{-1}-\overline{G}^{*-1} \big)$ \cite{kunn} as:
\begin{equation}\label{equa15}
\overline G(\varepsilon_{+}, {\bf p}_{+}) \overline G^{*}(\varepsilon_{-}, {\bf p}_{-}) = \frac{\overline{G}(\varepsilon_{+}, {\bf p}_{+})\ - 
\overline{G}^{*}(\varepsilon_{-}, {\bf p}_{-}) }{\varepsilon_{-} - \varepsilon_{+}- ({\bf p - \frac{1}{2}q})^{2} + ({\bf p + \frac{1}{2}q})^{2} - \Sigma^{*}(\varepsilon_{-}, {\bf p}^{-}) +\Sigma(\varepsilon_{+}, {\bf p}^{+})},
\end{equation}
which can be rewritten in  a compact form as:
\begin{equation}\label{equa5}
\overline G(\varepsilon_{+}, {\bf p}_{+}) \overline G^{*}(\varepsilon_{-}, {\bf p}_{-}) =
 \frac{\Delta G_{\bf p}(\varepsilon, \Omega,\ {\bf q})}{-\Omega + 2\,({\bf {p}  \cdot \bf{q}}) +\Delta \Sigma_{p}(\varepsilon, \Omega, {\bf q})},
\end{equation}
where 
$\Delta G_{\bf p}(\varepsilon, \Omega,\ {\bf q})=  \overline G(\varepsilon_{+}, {\bf p}_{+}) -  \overline G^{*}(\varepsilon_{-}, {\bf p}_{-})$, and
$\Delta \Sigma_{p}(\varepsilon, \Omega,\ {\bf q}) =  \Sigma(\varepsilon_{+}, {\bf p}_{+}) -  \Sigma^{*}(\varepsilon_{-}, {\bf p}_{-})$.\\
This leads to the standard quantum kinetic equation:
\begin{eqnarray}\label{equa05}
 \left[ -\Omega + 2 ( {\bf p\cdot q}) + \Delta \Sigma_{p}(\varepsilon, \Omega,\ {\bf q}) \right] \phi_{\bf p,p'}(\varepsilon, \Omega,\ {\bf q})
& = & \Delta G_{\bf p}(\varepsilon, \Omega,\ {\bf q})\delta_{\bf p,p'}   \nonumber \\ 
 & +  &  \Delta G_{\bf p}(\varepsilon, \Omega,\ {\bf q})\sum _{p''}  U_{\bf p,p''} (\Omega,\ {\bf q})\phi_{\bf p'',p'}(\varepsilon, \Omega,\ {\bf q}). 
\end{eqnarray}
As shown by Vollhardt and W\"olfle \cite{ref17}, the irreducible vertex $U_{\bf p,p'}$ 
and the self-enegy $\Sigma_{\bf p}$ are related with each other through the Ward identity 
\begin{equation}\label{equa7}
 \Delta \Sigma_{p}(\varepsilon, \Omega,\ {\bf q}) = \sum _{p'} \Delta G_{\bf p'}(\varepsilon, \Omega,\ {\bf q}) U_{\bf p', p} (\varepsilon, \Omega,\ {\bf q}),
\end{equation}
which describes the flux conservation.\\
Upon integrating over $p$, we obtain a simpler form of the continuity equation that describes the local conservation of the probability density
\begin{eqnarray}\label{equa25}
 -\Omega\ P_{\bf p'}(\varepsilon, \Omega,\ {\bf q})\ + {\bf q} \cdot  {\bf j} _{\bf p'} (\varepsilon, \Omega,\ {\bf q})\  \ 
 = \Delta G_{\bf p'}(\varepsilon, \Omega,\ {\bf q}),
 \end{eqnarray}
where $\Omega = i\partial_{t}$ and ${\bf q} = i{\bf \nabla}$.   
This conservation equation which is similar to that obtained in Ref.\cite{kunn} is valid for any initial distribution.

The probability density is written as: 
$P_{\bf p'}(\varepsilon, \Omega,\ {\bf q}) = \sum_{p} \phi_{\bf p,p'} (\varepsilon, \Omega,\ {\bf q})$ 
and the current density reads: ${\bf j}_{\bf p'} (\varepsilon, \Omega,\ {\bf q})= \sum_{p} 2 {\bf p} \cdot {\bf \phi}_{\bf p,p'} (\varepsilon, \Omega,\ {\bf q})$.\\
One can easily check from Eq.~(\ref{equa05}) that 
\begin{equation}\label{equa8}
\sum_{pp'} \phi_{\bf p,p'}(\varepsilon, \Omega,\ q=0) = \frac{\sum_{p} \Delta G_{\bf p}(\varepsilon, \Omega,\ q=0)}{-\Omega},
\end{equation}
where the factor $1/\Omega$ indicates that the total integrated density is conserved. 
The Ward identity (\ref{equa7}) implies that $L_{\varepsilon,\bf p}= \Delta G_{\bf p} (\varepsilon)$ is a good choice for the eigenfunction at least for large distance $q\rightarrow 0$ 
and long time, $\Omega \rightarrow 0$.
Therefore, we write
\begin{equation}\label{equa9}
L_{\varepsilon, \bf p}(\Omega=0, {\bf q})= \Delta G_{\bf p}(\varepsilon,{\bf q}) - \arrowvert G(\varepsilon,{\bf p})\rvert^{2}\Gamma({\bf p,q}) + O(q^{2}),
\end{equation}
where $\Gamma({\bf p,q})$ is supposed to be linear in $\bf{q}$ and  can in isotropic media be written as: $\sim {\bf j} ({\bf p}) 2({\bf p} \cdot {\bf q})$. \\
Replacing Eq.(\ref{equa2}) into Eq.(\ref{equa05}) for $\Omega =0$, one gets the following expression:
\begin{eqnarray}\label{equa050}
 \left[ 2 ( {\bf p\cdot q}) + \Delta \Sigma_{p}(\varepsilon, {\bf q}) 
 \right] \big[\Delta G_{\bf p}(\varepsilon, {\bf q})\ - \arrowvert G({\bf p},\varepsilon)\rvert^{2} {\bf j}(\varepsilon,{\bf p}) 2({\bf p} \cdot {\bf q}) \big] \ L_{\varepsilon,\bf p'}({\bf q})
& = &  D(\varepsilon) q^{2} \Delta G_{\bf p}(\varepsilon, {\bf q})\delta_{\bf p,p'}  \\ 
 & +  & \Delta G_{\bf p}(\varepsilon, {\bf q})\sum _{p''}  U_{\bf p,p''}(\varepsilon, {\bf q})
\big[\Delta G_{\bf p''}(\varepsilon, {\bf q}) \nonumber \\ 
 & - & \arrowvert G(\varepsilon,{\bf p''})\rvert^{2} {\bf j} (\varepsilon,{\bf p''})  2({\bf p''} \cdot {\bf q}) \big] L_{\varepsilon, \bf p'}( {\bf q}).\nonumber 
 \end{eqnarray}
Expanding the identity (\ref{equa7}) linearly in $q$ and inserting it into Eq.(\ref{equa050}), we obtain 
\begin{eqnarray}\label{equa051}
 \left[ 2 ( {\bf p\cdot q}) + \Delta \Sigma_{p}(\varepsilon, {\bf q}) 
 \right] \Delta G_{\bf p}(\varepsilon, {\bf q})\ - \Delta G_{\bf p}(\varepsilon, {\bf q}) 
 {\bf j}({\bf p})2({\bf p} \cdot {\bf q})
& = & \Delta G_{\bf p}(\varepsilon, {\bf q})\Delta \Sigma_{\bf p}(\varepsilon, {\bf q}) \\ 
 & -  & \Delta G_{\bf p} (\varepsilon, {\bf q}) \sum _{p''}  U_{\bf p,p''}(\varepsilon, {\bf q})
 \arrowvert G(\varepsilon,{\bf p''})\rvert^{2} {\bf j} (\varepsilon,{\bf p''})  2({\bf p''} \cdot {\bf q}). \nonumber 
\end{eqnarray} 
After some algebra, the quantum kinetic equation takes the form
\begin{equation} \label{equa555}
 2 ( {\bf p\cdot q}) + \sum _{p''}  U_{\bf p,p''}(\varepsilon,\ {\bf q})
 \arrowvert G(\varepsilon,{\bf p''})\rvert^{2} {\bf j} (\varepsilon,{\bf p''})  2({\bf p''} \cdot {\bf q}) = {\bf j}(\varepsilon, {\bf p})2({\bf p} \cdot {\bf q}).
\end{equation}
The current density is immediately found to be given by Eq.(\ref{equa555}) with ${\bf q}$ replaced by ${\bf p}$
\begin{eqnarray}\label{equa12}
{\bf j} (\varepsilon,{\bf p}) = 1 + \sum_{p''}\frac{{\bf p}''\ \cdot\ {\bf p}}{p^{2}} U_{\bf p,p''}(\varepsilon,0)
\arrowvert G(\varepsilon,{\bf p''})\rvert^{2} {\bf j} (\varepsilon,{\bf p''}).
\end{eqnarray}
This result is independent of the small momentum ${\bf q}$ since in deriving Eqs. (\ref{equa051})-(\ref{equa12}) we already kept 
only linear terms in ${\bf q}$, we therefore put ${\bf q}=0$ in $U_{\bf p,p''}(\varepsilon,\ {\bf q})$ \cite{ref17,VW,ref18}.
Equation (\ref{equa12}) clearly shows that in lowest-order perturbation theory, $j (\varepsilon,{\bf p}) = 1$.
The term $({\bf p}'' \cdot{\bf p}/p^{2}) U_{\bf p,p''}$ defines  the  cross section of all interference contributions in multiple scattering.  
Analytical solutions of Eq.(\ref{equa12}) for 3D speckle potentials are shown in Appendix.

In order to find the eingenfunction of the Bethe-Salpeter equation,  we employ the following transformation \cite{ref18}:
\begin{equation}\label{equa013}
 \Delta G_{\bf p}(\varepsilon,{\bf q}) = G(\varepsilon, {\bf p + q}/2)\ - G^{*}(\varepsilon, {\bf p -q}/2)=2i \text {Im} G_{\bf p} (\varepsilon)\ + 
 \ \frac{\partial \text{Re}\,G_{\bf p}(\varepsilon)}{\partial {\bf p}} \cdot  {\bf q}.
\end{equation}
Using the fact that $$\frac{\partial \text{Re}\,G_{\bf p}(\varepsilon)}{\partial {\bf p}} \cdot {\bf q}= 2\,{\bf p} \cdot {\bf q}\frac{\partial \text{Re}\,G_{\bf p}(\varepsilon)}{\partial p^{2}},$$ 
then introducing Eq.(\ref{equa013}) into  Eq.~(\ref{equa9}), one obtains for the eigenfunction of the Bethe-Salpeter equation 
\begin{eqnarray}\label{equa13}
L_{\varepsilon, \bf p} ({\bf q}) = 2i \,\text {Im}G_{\bf p}(\varepsilon)\ + 2 \ {\bf p} \cdot {\bf q} \frac{\partial \text{Re}\,G_{\bf p}(\varepsilon)}{\partial p^{2}} - 
\ \arrowvert G(\varepsilon,{\bf p})\rvert^{2} 2\ ({\bf p} \cdot {\bf q}) {\bf j} (\varepsilon, {\bf p}),
\end{eqnarray}
which admits an exact solution in linear order of $q$. 

The spectrale function $\text {Im} G_{\bf p}(\varepsilon)$ contains all the information about the diffusion of the particles in  a disordered medium. It is given by
\begin{equation}\label{A}
 -\text {Im}G_{\bf p} (\varepsilon)= \frac{\text {Im}\Sigma_{\bf p}(\varepsilon)}{\arrowvert {\varepsilon - \varepsilon_{\bf p}\ -\Sigma_{\bf p}(\varepsilon)}\rvert^{2}},
\end{equation}
where the self-energy has the following form $\Sigma_{\bf p}(\varepsilon) = \sum_{\bf p'}\ U_{\bf p,p'}[\varepsilon-\varepsilon_{\bf p'}- \Sigma_{\bf p'}(\varepsilon)]^{-1}$ which 
is a negative function (represents the energy width of the spectral function). Thus,
\begin{equation}\label{B}
\arrowvert G(\varepsilon,{\bf p})\rvert^{2} = \frac{\text {Im} G_{\bf p}(\varepsilon)} {\text {Im} \Sigma_{\bf p}(\varepsilon)}.
\end{equation}
The scattering mean-free time is defined as:
\begin{equation}\label{C}
\tau_{\bf p} (\varepsilon) = -\frac{\hbar} {2 \ \text {Im} \Sigma_{\bf p}(\varepsilon)}.
\end{equation}
Upon inserting  Eqs.(\ref{A})-(\ref{C}) into Eq.(\ref{equa13}) and removing $-2i$, one finds:
\begin{equation}\label{equa14}
L_{\varepsilon, \bf p} ({\bf q}) = - \text {Im}G_{\bf p} (\varepsilon)\ - i ({\bf p}\cdot {\bf q})\ \gamma({\bf p},\varepsilon), 
\end{equation}
where
\begin{equation}\label{G}
\gamma (\varepsilon,{\bf p})= \ \text {Im}G_{\bf p} (\varepsilon)\, 2\tau_{\bf p} (\varepsilon) {\bf j} (\varepsilon,{\bf p})\ - \frac{\partial \text{Re}\,G_{\bf p} (\varepsilon)} {\partial p^{2}}\ +O(q^{2}),
\end{equation}
is called  the dipolar contribution which generates the flow of energy and thus, provides corrections that support a current density.
The solution of the Bethe Salpeter equation $L_{\varepsilon,\bf p}({\bf q})$ at small $q$ is governed by the spectral function  ($A(\varepsilon,{\bf p}) = - \text{Im}G_{\bf p}(\varepsilon)/\pi$). 

Finally, the normalization constant $K$ can be determined by matching  Eq.~(\ref{equa8}) and Eq.(\ref{equa2}) for $q=0$.
This yields
\begin{equation}\label{equa20}
 K \sum_{pp'} \frac{\text{Im}G_{\bf p}(\varepsilon) \text{Im}G_{\bf p'}(\varepsilon)}{-i\Omega} = \frac{ \sum\limits_{p}2i\text{Im}G_{\bf p}(\varepsilon)}{-\Omega},
\end{equation}
we thus, immediatly deduce that 
\begin{equation}\label{equa20a}
K= \frac{2}{\sum\limits_{p} - \text{Im} G_{\bf p}(\varepsilon)}.
\end{equation}
Replacing the resulting constant  (\ref{equa20a}) into Eq.(\ref{equa2}), one gets for the quantum propability 
\begin{equation}\label{equa21}
 \phi_{\bf p,p'}(\Omega,\ {\bf q}) = \frac{2}{\sum\limits_{p} -\text {Im}G_{{\bf p}}(\varepsilon)}  \frac{L_{\varepsilon, \bf p}({\bf q}) L_{\varepsilon, \bf p'}({\bf q}) }{-i \Omega \ + \ D q^{2}},
\end{equation}
which has a diffusion pole according to \cite{kunn,Piraud,ref18}.
The sum $\sum\limits_{p} -\text {Im}G_{\bf p}(\varepsilon)$ is linked to the density of states  per unit volume $\rho(\varepsilon)$ as: 
$\sum\limits_{p} -\text {Im}G_{\bf p} (\varepsilon)= \pi \rho(\varepsilon) $.\\
For $p\longrightarrow \infty$, $\text{Im}\Sigma_{\bf p}(\varepsilon)$ vanishes, we then expect that $j(p)\longrightarrow 1$.
The dipolar contribution  $\gamma(\varepsilon,{\bf p})$ has the following form
\begin{equation}\label{equa22}
\gamma_{0} (\varepsilon,{\bf p}) =  \ \ \arrowvert G(\varepsilon,{\bf p})\rvert^{2}\ - \frac{\partial \text{Re}\,G_{\bf p}(\varepsilon)}{\partial p^{2}}.
\end{equation}
Here $\gamma_{0}  (\varepsilon,{\bf p})$ describes multiple scattering as a sequence of scattering events where both retarded and advanced amplitudes travel along the same path.
When $p\longrightarrow \infty$,  the kinetic energy of the atoms exceeds the exitation potential, a large isotropic waves travel without feeling the disorder effect \cite{ref15,kunn}.


\subsection{Diffusion constant from quantum localization corrections for isotropic scattering} \label{DC}

We now concentrate on the calculation of the diffusion constant using the quantum kinetic equation derived in the previous section.
Let us write the expression of the current density 
\begin{equation}
 {\bf j}_{\bf p'}(\varepsilon, \Omega,\ {\bf q}) = 2 \sum _{p}\ {\bf {p} \cdot \phi_{\bf p,p'}}(\varepsilon, \Omega,\ {\bf q}).
\end{equation}
Incorporating  $\phi_{\bf p,p'}$ from Eq.(\ref{equa21})  into  $ {\bf j}_{\bf p'}(\varepsilon, \Omega,\ {\bf q})$, we find 
\begin{equation}\label{z}
 {\bf j}_{\bf p'}(\varepsilon, \Omega,\ {\bf q})= \frac{2}{\pi \rho(\varepsilon)}\frac{L_{\varepsilon, \bf p'}({\bf q})}{-i \Omega\ + \ D q^{2}} \sum _{p} 2 {\bf p} \ L_{\varepsilon, \bf p}({\bf q}).
\end{equation}
Inserting the corrections manifesting in $L_{\varepsilon, \bf p}({\bf q})$ into Eq.~(\ref{z}), we obtain
\begin{equation}\label{27}
 {\bf j}_{\bf p'}(\varepsilon, \Omega,\ {\bf q}) = \frac{2}{\pi \rho(\varepsilon) }\frac{L_{\varepsilon, \bf p'}( {\bf q}) } 
{-i \Omega + \ D q^{2}} i\ {\bf q}\ \ \frac{2}{3} \sum _{p} \bigg[- \ \text {Im}G_{\bf p}(\varepsilon) 2\tau_{\bf p} (\varepsilon) {\bf j}(\varepsilon,{\bf p}) p^{2}\ - \ p^{2} \frac{\partial \text{Re}\, G_{\bf p}(\varepsilon)}{\partial p^{2}} \bigg].
\end{equation}
The factor $(1/3)$ that appears in Eq.~(\ref{27}) comes from the angular integral.

The probability density  is defined as: 
\begin{equation}
P_{\bf p'}(\varepsilon, \Omega,\ {\bf q})=\sum_{p} \phi_{\bf p,p'} (\varepsilon, \Omega,\ {\bf q})= 2\ \frac{L_{\varepsilon, \bf p'}({\bf q}) }{-i \Omega \ + \ D q^{2}}.
\end{equation}
According to Fick’s law, which relates the diffusive flux to the concentration gradient in the diffusive regime ($\Omega, {\bf q} \rightarrow 0$),
$ {\bf j}_{\bf p'}(\varepsilon, \Omega,\ {\bf q}) = i\ {\bf q} D(\varepsilon)  P_{\bf p'} (\varepsilon, \Omega,\ {\bf q})$,
the diffusion constant is extracted as: 
\begin{equation}\label{equa20}
 D(\varepsilon)\ = \ \frac{1}{\pi\rho(\varepsilon)} \frac{2}{3} \sum_{p}p^{2} 
 \left( - \text {Im}G_{\bf p}(\varepsilon) 2\tau_{\bf p} (\varepsilon)  {\bf j} (\varepsilon,{\bf p})\ - \ p^{2}\frac{\partial \text{Re}\,G_{\bf p}(\varepsilon)}{\partial p^{2}}\right).
\end{equation}
This equation is appealing since it includes weak localization corrections.
In the regime of a weak disorder, $\text {Im} G_{\bf p}(\varepsilon)$  is strongly peaked near $p= \sqrt\varepsilon$, yielding $\pi \rho(\varepsilon)\approx \sqrt{\varepsilon}/(4\pi)$.
Therefore, $D(\varepsilon)$ reduces to:
\begin{equation}\label{equa22}
 D(\varepsilon)=  \frac{4\pi}{\sqrt{\varepsilon}} \frac{2}{3}\frac{4\pi}{(2\pi)^{3}} \pi \int_{0}^{\infty} dp p^{4}\delta(\varepsilon-p^{2}) 2\tau_{\bf p} (\varepsilon) {\bf j}({\varepsilon,\bf p}).
\end{equation}
Keeping in mind that the second term is negligible, we find:
$D(\varepsilon)=  4 \varepsilon\, j(\sqrt{\varepsilon})\tau_{p=\sqrt{\varepsilon}}/3$.
Asuming that $j(\sqrt{\varepsilon})=1$ at large $\varepsilon$, we recover the familiar expression 
for classical wave namely $ D(\varepsilon) =  v^2\tau^{*}/3$ \cite{Shap,Boudj},
where $v$ is the wave velocity, and $\tau^{*}$ is a transport mean free time  which is
straightforwardly associated with the transport mean path $\ell^{*}$ via $\tau^{*}= \ell^{*}/v$.

In the next section, we will discuss the critical regime of transition by numerically solving the above Bethe-Salper equation.

\section{Numerical results} \label{NR}

In this section, we solve numerically the Bethe-Salpeter equation for the transport phenomenon of BEC subjected to a laser speckle potential. 
To do so, we apply the self-consistent Born approximation to calculate the self-energy which is the key factor of the diffusion.
The current density $j(p)/p$ is computed iteratively, and then insert the result into Eq.(\ref{equa20}) to obtain the diffusion coefficient.
The number of required iterations for convergence is $100$ iterations which are enough to obtain the desired solution \cite{Boudj}.

From now on, all energies, including $\Sigma$, are expressed in units of the  quantum correlation energy  $\varepsilon_{\sigma}=1/(2m\sigma^2)$ with
$\sigma$ being the correlation length, momenta are expressed as $p =k$, where the de-Broglie wave number $k$ is scaled in units of $1/\sigma$  (we recall that $\hbar$=1).
We choose a constant value of the disorder amplitude $(V_{R}=0.5\varepsilon_{\sigma})$ 
for which the condition  of  a perturbative disorder to be valid $ \varepsilon_{k}\gg V_{R}^{2}/\varepsilon_{\sigma}$ \cite{delande}. 

\begin{figure}\label{jp}
\begin{center}
\includegraphics[angle=360,width=0.7\columnwidth] {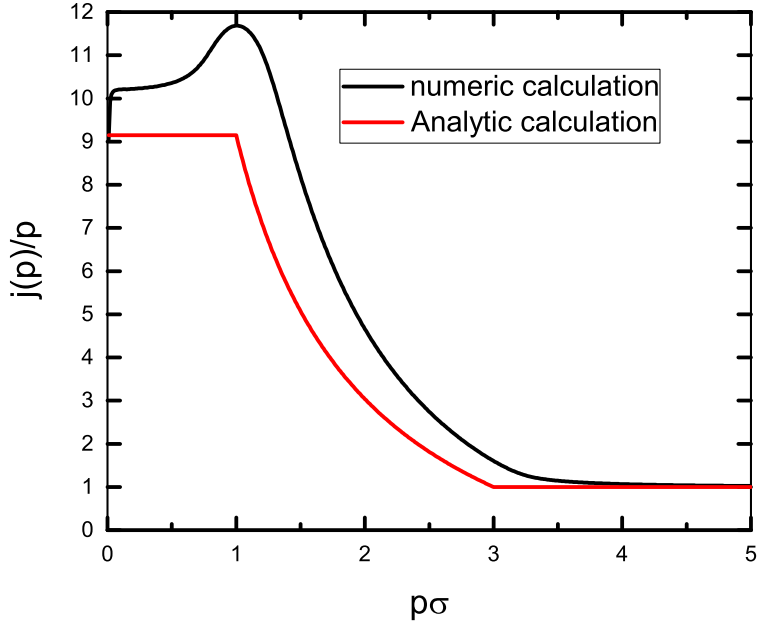}
\caption{Static current density $ j(p)/p $ as a function of $p\sigma$ for $\varepsilon_{c}= 0$. }
\label{jp.eps}
\end{center}
\end{figure}

In Figure \ref{jp.eps}\label{jp} we report the numerical and analytical results of  the current density $j(p)/p$ as a function of the scaled wave number $p\sigma$.
We see that the current density reaches its maximum at $p\sigma \lesssim 1$, then it decreases and becomes constant for large momenta.  
The analytical results and the numerical simulation diverge from each other notably for small momenta.  
A closer look at the same figure shows that the numerical simulation predicts a hump in the current density at $p\sigma \sim 1$, indicating that more atoms can participate in the multiple scattering. 
These atoms are dephazing from their initial positions most probably due to the effect of a weak localization. 
Both methods  show that the current diffusion converge to $1$ at very large momentum $(p\rightarrow \infty)$.
In such a regime weak localization corrections disappear and a classical diffusion takes place (see also  Eq.~(\ref{equa12})).

\begin{figure}
\begin{center}
\includegraphics[angle=360,width=0.7\columnwidth]{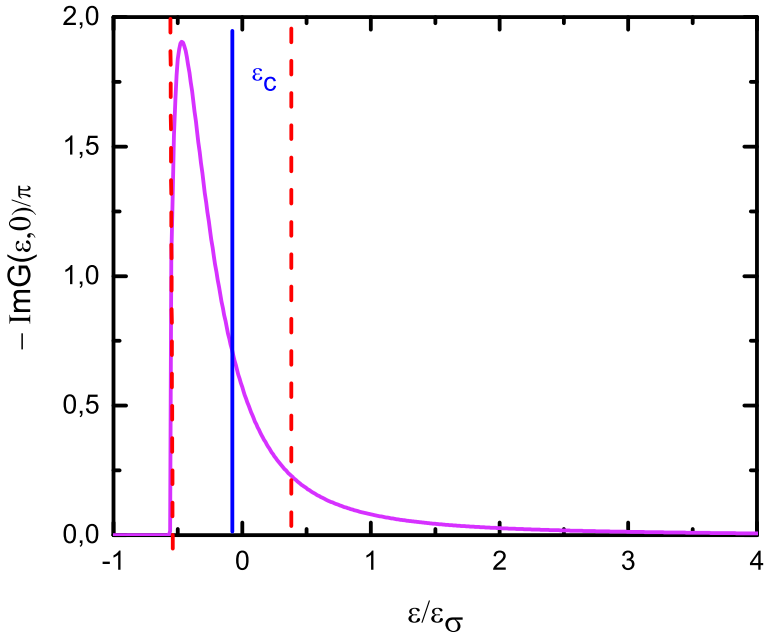}
\caption{Spectral function for $p= 0 $ and $V_{R} = 0.5 \varepsilon_{\sigma}$.}
\label{aex.eps}
\end{center}
\end{figure}

To calculate the spectral function, we solve numerically  Eq.~(\ref{A}).  The results are shown in Figure \ref{aex.eps}\label{aex}. 
We observe the appearance of  a peak at $\varepsilon= - 0.5\varepsilon_{\sigma}$. The width half height of this peak $\Delta \varepsilon= 0.4\varepsilon_{\sigma}$  is 
inversely proportional to the mean free time of the atoms in speckle.
At low energies, the relevant parameter of localization is the typical length $\ell_{c}$ which is associated with a typical momentum $k_{c} =\sqrt{2 m \varepsilon_c}$.
Note that the width of the spectral function at $\varepsilon = \varepsilon_{c} $ is of the order of $\ell_{c} \sim \sigma /(V_{R}/\varepsilon_{\sigma})^{2}$\cite{ref23}.
In our case, we find $\ell_{c} \approx 4 \sigma$. 
From the same figure we see that the mobility edge is located at  $\varepsilon_{c}/\varepsilon_{\sigma}=- 0.04$  which implies that $k_{c}/k_{\sigma}=0.2$. 
In the limit of weak disorder,  we find $k_{c}\ell_{c}=0.8$ satisfying the celebrated Ioffe-Regel criterion namely, $(k_{c}\ell_{c}\sim 1)$ \cite{BART}. 
For $\ell_{c} \gg  4 \sigma$, the density atomic exceeds the optical potential which means  that $\ell_{c}$ varies faster than the modulations of the speckle $\sigma$.
In such a situation we fall to a classical diffusion.

\begin{figure}
\begin{center}
\includegraphics[angle=360,width=0.7\columnwidth]{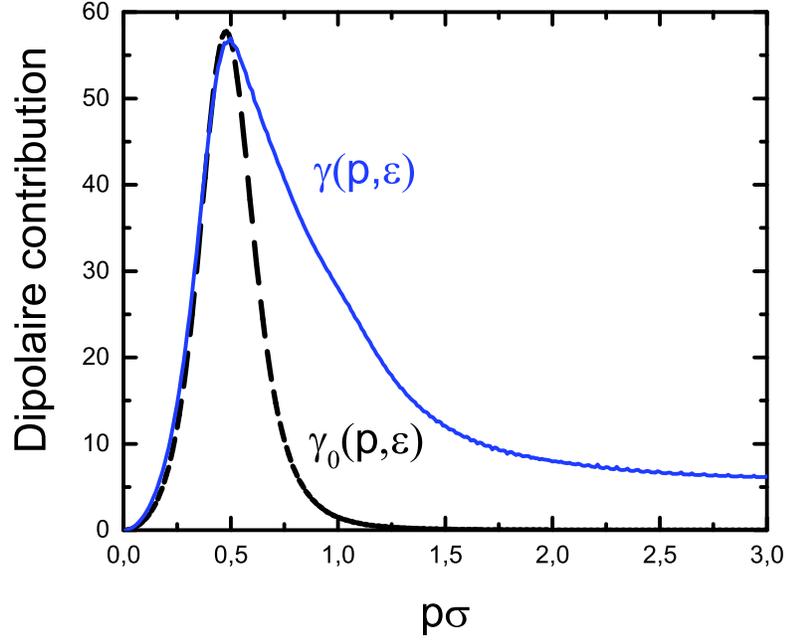}
\caption{ Dipolar contributions $\gamma (p,\varepsilon) $ and $\gamma_0 (p,\varepsilon) $ for $\varepsilon/\varepsilon_{\sigma}=-0.1$.
Here $\gamma (p,\varepsilon) $ is associated with coherent matter wave transport (in the presence of the quantum corrections)
and $\gamma_0 (p,\varepsilon) $ corresponds to the incoherent matter wave transport (in the absence of the quantum corrections).}
\label{BS.eps}
\end{center}
\end{figure}

The solution of Bethe-Salpeter equation for the dipolar contribution in the regime of coherent matter wave transport (i.e. in the presence of the quantum corrections, $\gamma (\varepsilon,p)$) 
and for the incoherent matter wave transport (i.e. classical regime, $\gamma_0 (p,\varepsilon) $ is captured in figure \ref{BS.eps}\label{BS}. 
We see that the two curves are almost indistinguishable, yielding an excellent agreement in the region $p\sigma < 0.5$.
Whereas for $p\sigma> 0.5$, the curves diverge form each other (the width of quantum particles curve is broadened) due
to the enhancement of the correlations induced by weak localization. The narrow spectrum in the classical regime indicates the absence of quantum correlation effects.
In the presence of the quantum corrections, the energetic  atoms are shifted by the interference giving rise to increase  the intensity of the dipolar contribution
results in a large anisotropic momentum because of their deviations by the disorder effect \cite{Piraud}.
Given that  $p\sigma$  equivalent to $ \upsilon_{p}/\upsilon_{\sigma}$, where $v_{\sigma}=1/\ m\sigma$, we find that
the dipolar contribution $ \gamma(\varepsilon, p)$ is dominated by atoms with velocity  $v_{\sigma} = 0.66 / m\sigma$.
We detect on the other hand  a fraction of atoms faster than $v_{\sigma} = 1.11 / m\sigma$ 
contributing  to the multiple diffusion. Our results show a large spectrum in $ \gamma(\varepsilon, p)$ compared to that obtained in Ref.\cite{ref18} 
since we take into account higher-order contribution to the current density $j(p)$.

\begin{figure}
\begin{center}
\includegraphics[angle=360,width=0.7\columnwidth]{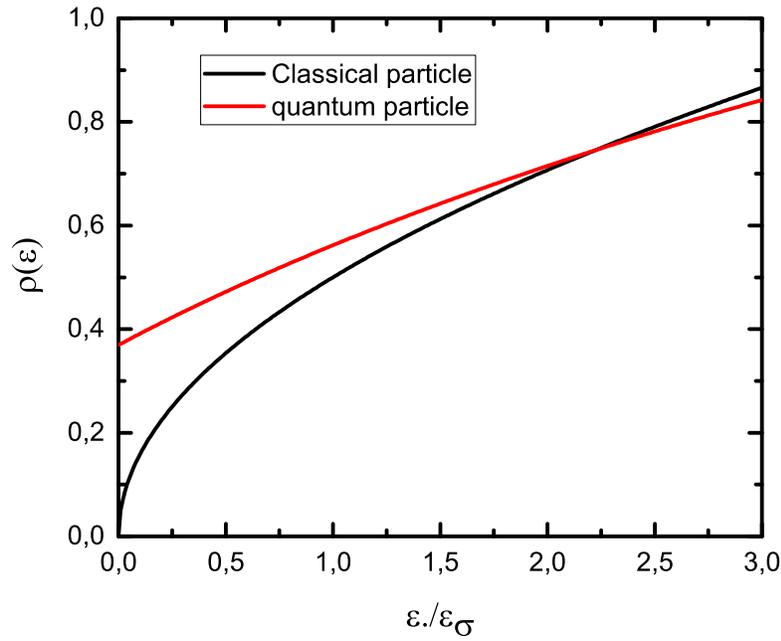}
\caption{The density of states per unit of volume as a function of $\varepsilon/\varepsilon_{\sigma}$ for $p= 0 $.
In the classical diffusion the density of states is given as: $ \rho(\varepsilon)  = 4\pi\delta(\varepsilon - \varepsilon_{p})$.}
\label{DOS.eps}
\end{center}
\end{figure}

These quantum corrections have an important impact on the distribution of waves packet. 
This can be seen in the density of states $\rho(\varepsilon) = \int \text{ Im} G(\varepsilon,p) d {\bf p}/(2\pi)^3$
which measures the average number of states in the random medium per unit of volume.
Figure \ref{DOS.eps}\label{DOS} depicts that at energies $\varepsilon \ll \varepsilon_{\sigma}$, the probability is small but finite.
Our numerical calculation predicts that around $38\%$ of atoms are localized due to the interference effects. 
Here the fraction of localized atoms is given by $f_{\text{loc}}\approx\int_{-\infty}^{\varepsilon_{c}} \ d\varepsilon \text{ Im} G(\varepsilon,p=0)$ \cite{ref23}.
At higher energies,  the probability grows with the energy. 
In such a situation, one has $\text {Im} G(\varepsilon,p)= -\pi \delta(\varepsilon - \varepsilon_{p})$, hence the atoms do not feel the disorder potential and the system attains the classical diffusion. 
Consequently, the density of states behaves like the free-space expression $\rho(\varepsilon) \sim \varepsilon^{1/2}$, where the dispersion relation reads
$\varepsilon= p^{2}/2m$ (see black line in figure \ref{DOS.eps}\label{DOS}).
It is clear that the two curves (quantum and classical) cross with each other at $\varepsilon \simeq 2.25 \varepsilon_{\sigma}$
pointing out that the atoms diffuse in a classical way at energies $\varepsilon \gtrsim 2.25 \varepsilon_{\sigma}$.

\begin{figure}\label{DU}
\begin{center}
\includegraphics[angle=360,width=0.7\columnwidth]{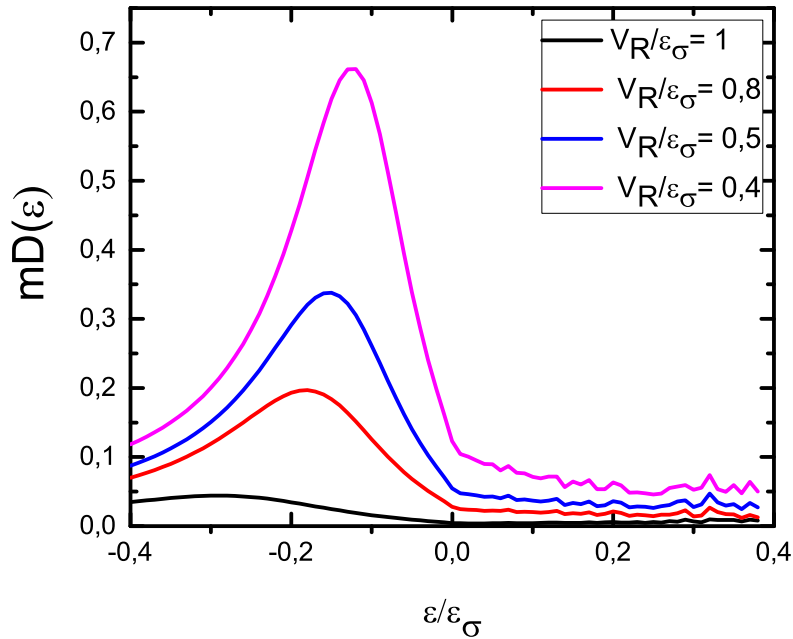}
\caption{Diffusion coefficient as a function of the energy for several values of the disorder amplitude and for $p= 0 $.}
\label{DU.eps}
\end{center}
\end{figure}

In Figure \ref{DU.eps}\label{DU}, we plot the diffusion constant  as a function of the energy  for different  disorder amplitudes.
Our results reveal that the diffusion coefficient $mD(\varepsilon)$ reaches its maximum at negative energies for any disorder amplitude, then 
it decreases when the energy becomes close to zero. 
Such a decay in the diffusion coefficient which depends also on the scattering amplitude as is seen in figure \ref{DU.eps}\label{DU},
is due to the weak localization  effects induced by the fluctuations caused by the number of scatterers.

\begin{figure}\label{DE}
\begin{center}
\includegraphics[angle=360,width=0.7\columnwidth]{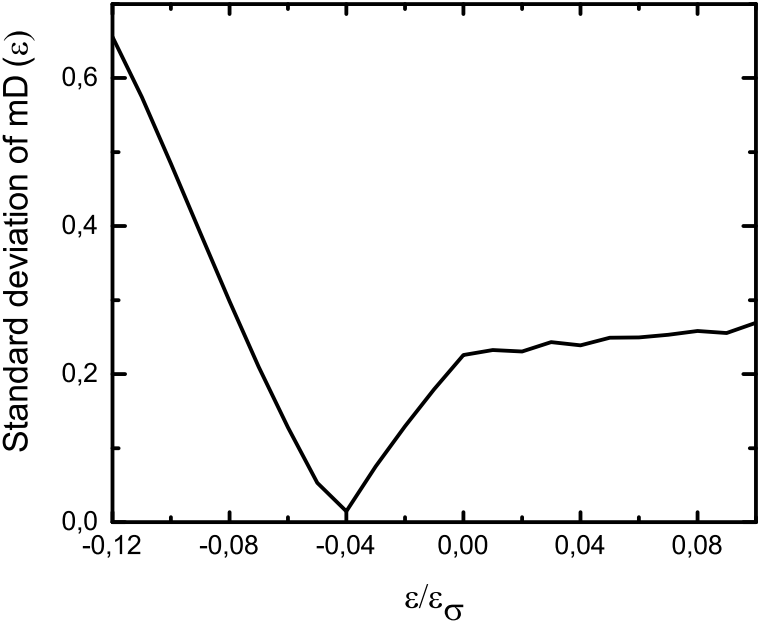}
\caption{Standard deviation of the diffusion coefficient in a critical regime for $V_{R}/\varepsilon_{\sigma}=0.5$.}
\label{DE.eps}
\end{center}
\end{figure}

Let us discuss more profoundly the transport  properties from diffusion coefficient.
To this end, we use the standard deviation to approach the transition point.
Figure \ref{DE.eps}\label{DE} shows three distinct regions.
In the first region,  $-0.12<\varepsilon/\varepsilon_{\sigma}<-0.04$, $m D(\varepsilon)$ decreases rapidly with the energy until it attains the mobilty edge
$\varepsilon_c =-0.04 \varepsilon_{\sigma}$.
The situation is different in the second region, $-0.04<\varepsilon/\varepsilon_{\sigma}<0$, where the diffusion coefficient  increases linearly.
A similar behavior holds true for a Gaussian disorder potential \cite{ref23}. In this region,  the states cease to extend over the disorder potential and thus delocalize.
In the third interval $\varepsilon>0$ and for small energies,  $m D (\varepsilon)$ is saturated to a constante value ($m D (\varepsilon)\approx 0.25$) 
owing to the weak localization effects. 
This phenomenon is important because the wave transport is quite sensitive to large momentum  which leads to increase  the interferences between scattered atoms
evoking a deviation of fast atoms from its initial position and hence reducing the constant diffusion.
This is in stark contrast with the Dirac-peaks potential \cite{ref24}, where the atoms are insensitive to the disorder effects.
Note that in the anisotropic case, the diffusion coefficient increases in power law in the limit of low energies while it behaves as $D\propto \varepsilon^{5/2}$ at higher energies \cite{Piraud}.

\section{Conclusion} \label{Conc}

In this paper, we examined  the transport and localization of matter waves in isotropic 3D speckle potentials using the Bethe-Salper equation and the self-consistent theory of
Anderson localization.
We calculated in particular the fundamental transport quantities such as the current intensity, the dipolar contribution, the density of states, the spectral function, 
and the reduced diffusion constant.
We found that these quantities deviate from their classical counterpart due to corrections arising from multiple scattering of matter waves. 
At low energy, our numerical calculations predict that the weak localization corrections increase the probability density and may shift the dipolar contribution.
The reduction of the diffusion constant which depends on the scattering amplitude signals the occurrence of weak localization effects.
Our results pointed out also  that the diffusion process continues with the initially fast atoms until the transport means path becomes of the order of the wavelength of the condensate.

\section*{CRediT author statement}
Afifa Yedjour: Conceptualization, Methodology, Software, Data curation, Writing- Original draft preparation. \\
Abdelaali Boudjemaa: Visualization, Investigation, Writing-Review and Editing

\section*{Acknowledgments}

AY gratefully acknowledges the helpful discussions with Bart Van Tiggelen. 

\newpage 

\section{Appendix}

In this appendix we derive an explicit expression for the density current in the case of a 3D laser speckle potential produced by diffraction 
which is widely used with quantum gases experiments, its correlation function is given by
$\langle U({\bf R})\rangle = V_R^2\mathrm{sinc}^2 (R/\sigma)$, where $V_R$ is the amplitude disorder. 
For $V_R \rightarrow \infty$ and $\sigma \rightarrow 0$, the speckle potential reduces to the uncorrelated white-noise random potential.
In Fourier space one can write:
\begin{equation}\label{equa2.}
U_{\bf p,p'} =  V_R^2 \int d{\bf R} \sum_{\bf p'} 
\dfrac{\sin^{2}\left(R/\sigma\right)}{\left(R/\sigma\right)^{2}} e^{i({\bf p}-\ {\bf p}') \cdot {\bf R}}.
\end{equation}
Inserting  $U_{\bf p,p'}$ into Eq.~(\ref{equa12}) and using the fact that $\sum_\mathbf{p'} \equiv \int d{\bf p'}/(2\pi)^3$, one finds
\begin{equation}\label{a2}
 j(\varepsilon,{\bf p}) = 1 + \frac{V_R^2}{(2\pi)^3} I({\bf p,\ p'}) \arrowvert G(\varepsilon,{\bf p'})\rvert^{2}j(\varepsilon,{\bf p'}),
\end{equation}
where
\begin{equation}\label{a1}
 I({\bf p,\ p'}) =\int d{\bf R} \int d{\bf p}' \frac{ {\bf p} \cdot \ {\bf p}'}{p^{2}} \dfrac{\sin^{2}\left(R/\sigma\right)}{\left(R/\sigma\right)^{2}}
 e^{i({\bf p}-\ {\bf p}') \cdot {\bf R}}.
\end{equation}
Setting  $\int d \hat{p''}=\ d p''  p''^{2}$, then the integral in momentum space becomes 
\begin{eqnarray}
\int\frac{d \hat{p'}}{4\pi} ({\bf p} \ \cdot\ {\bf p}') e^{i( {\bf p}-\ {\bf p}') \cdot {\bf R}}
&  =  & {\bf p} \cdot \frac{1}{p'} e^{\ i\ {\bf p}\cdot {\bf R}}\cdot
i \frac{\partial }{\partial {\bf R}} \int d \hat{p'} e^{-\ i\ {\bf p}''\cdot {\bf R}}  \\
&  =   &{\bf p} \cdot \frac{e^{ i\ {\bf p}'\cdot {\bf R}}}{p'} 
 i \frac{\partial }{\partial {\bf R}} \mathrm{sinc}( p'\ R)  \nonumber \\
&  =   & \frac{\hat{p}. \hat{R}}{R} e^{ i\ {\bf p}\cdot {\bf R}} i \frac{\partial}{\partial p'} \mathrm{sinc} (p' \ R). \nonumber
 \end{eqnarray}
Now the integral over $R$ can be evaluated as: 
 \begin{eqnarray}
  \int\frac{ d \hat{R}}{4\pi} \ \hat{R}\ e^{i{\bf p} \cdot {\bf R}} & = &
  \frac{1}{R}\frac{1}{i}\frac{\partial}{\partial\ p}\int \frac{d^{3} \hat{R}}{4\pi}
  e^{\ i\ {\bf p}\cdot {\bf R}} = \frac{\hat{p}}{R}\frac{1}{i}\frac{\partial}{\partial\ p} \mathrm{sinc} ( p\ R).
 \end{eqnarray}
With this, Eq~(\ref{a1}) can be rewritten as:
\begin{equation}\label{a4}
 I(p,\ p') =  4\pi \frac{\partial}{\partial\ p}\frac{\partial}{\partial\ p'} \int_{0}^{\infty}
 d R \,\mathrm{sinc}^{2}\left(\dfrac{R}{\sigma}\right) \mathrm{sinc} ( p\ R) \ \mathrm{sinc} ( p'\ R) =  4\pi \frac{\partial}{\partial\ p}\frac{\partial}{\partial\ p'} \frac{1}{pp'}\ K(p,p'),
\end{equation}
where
\begin{equation} \label{a5}
 K(p,p')= \int_{0}^{\infty}d R \ \mathrm{sinc}^{2} \left(\dfrac{R}{\sigma}\right) \left(\frac{\sin p'R}{R}\right) \left(\frac{\sin R}{R}\right).
\end{equation}
Putting
\begin{equation} \label{a6}
M(p,p') = \frac{\partial}{\partial\ p}\ \frac{\partial}{\partial\ p''}\ K(p,p')= \int_{0}^{\infty}\ d R\,\mathrm{sinc}^{2} \big(R/\sigma \big) \cos(p\ R)\ \cos(p''\ R).
\end{equation}
Then employing the relation $ \int_{-\infty}^{+\infty} dR e^{iqR}/R = \pi\Theta(q)$, we obtain:
\begin{equation} \label{a7}
  M(p,p') \ = \ \frac{\pi\sigma}{16}\left[ 2 \ - \ p \ - \ p'\ - \ 2 \lvert p\ - \ p'\rvert 
 \ + \lvert p\ + \ p' \ - \ 2\rvert\ \ + \  \lvert 2\ - \ p \ + \ p' \rvert\ 
\ + \ \lvert 2\ + \ p \ - \ p' \rvert\ \right].
\end{equation}
Upon substituting Eq.(\ref{a7}) into Eq.(\ref{a4}), we find
\begin{equation} 
  I(p,p') \ = \ \frac{4\pi^{2}\sigma}{16}\left[I_{1}\ + \ I_{2}\ + \ I_{3}\ + \ I_{4} \ \right],
\end{equation}
where 
$$I_{1} = 2\ \frac{p\ + \ p'\ - \ 2}{p^{2}p'^{2}}\left( - p^{2}\ - \ p'^{2}\ + \ p p'\ + 2\ + p\ + \ p' \ \right)
\left(\Theta\left(p \ + \ p'\ - \ 2\right)\ - \ 1\right),$$
 $$I_{2} = \ - \ 4\ \frac{p\ - \ p'\ }{p^{2}p'^{2}}\left(  p^{2}\ + \ p'^{2}\ + \ p p'\ \right) 
 \left(\left(\Theta(p \ - \ p'\ \right)\ + \ 1\right)\left(1 \ - \, \Theta\left(p' \ - \ p\ \right)\right),$$
 $$I_{3} = 2\ \frac{p'\ - \ p\ + \ 2}{p^{2}p'^{2}}\left(  p^{2}\ + \ p'^{2}\ + \ p p'\ - 2\ - p\ + \ p' \ \right)  \left(\Theta\left(p' \ - \ p\ + \ 2\right)\ - \ 1\right),$$
 $$I_{4} = 2\ \frac{p'\ - \ p\ - \ 2}{p^{2}p'^{2}}\left(  p^{2}\ + \ p'^{2}\ + \ p p'\ - 2\ + p\ + \ p' \ \right) \left(\Theta\left(p' \ - \ p\ - \ 2\right)\ - \ 1\right).$$
Equation (\ref{equa12})  was solved self-consistently in section \ref{NR}. Its behavior has been displayed in Fig.\ref{jp.eps}.

\end{document}